\documentclass[a4paper]{jpconf}
\usepackage{graphicx}
\usepackage{amssymb}
\usepackage{hyperref}
\usepackage{graphicx}
\usepackage{amsmath}
\usepackage{amssymb,amsfonts,amsthm}
\usepackage{cite}

\begin{document}
\title{Impact of new physics on $B+L$ violation at colliders}

\author{David G Cerde\~no$^1$, Peter Reimitz$^2$, Kazuki Sakurai$^3$ and Carlos Tamarit$^4$}
\address{$^1$ Institute for Particle Physics Phenomenology, Department of Physics,
Durham University, Durham DH1 3LE, United Kingdom}
\address{$^2$ Institut f\"ur Theoretische Physik, Universit\"at Heidelberg
Philosophenweg 16, 69120 Heidelberg, Germany}
\address{$^3$ Institute of Theoretical Physics, Faculty of Physics,
University of Warsaw, ul. Pasteura 5, PL02093 Warsaw, Poland}
\address{$^4$ Physik-Department T70,  Technische Universit\"at M\"unchen, James-Franck Stra{\ss}e 1, 85748 Garching, Germany}
\ead{$^1$davidg.cerdeno@gmail.com, $^2$p.reimitz@thphys.uni-heidelberg.de,\\ $^3$kazuki.sakurai@fuw.edu.pl, $^4$carlos.tamarit@tum.de}

\begin{abstract}
In the Standard Model, chiral electroweak anomalies predict nonperturbative interactions that violate baryon ($B$) plus lepton number ($L$). The
potential observability of these processes at colliders has been amply discussed in the literature, mostly focusing on the impact of the accompanying boson emission, which contributes
to the cross sections through an exponential function of the center-of-mass energy. We focus instead on the impact of exotic fermions charged under $SU(2)_L$, which not only can be emitted in these processes, but also affect the non-exponential contributions to the cross-sections. Estimating the latter using instanton techniques, we find sizable effects that suggest that if $B+L$-violating processes are ever seen at colliders, they may involve physics beyond the Standard Model.
\end{abstract}

\section{Introduction}
Processes that violate baryon number are of extreme interest in particle physics, as baryon violation is essential for explaining the abundance of matter over anti-matter in the Universe. In the Standard Model $B$ and $L$ are accidental symmetries, and are in fact violated by non-perturbative sphaleron processes \cite{Manton:1983nd,Dashen:1974ck,Klinkhamer:1984di}, which conserve $B-L$. Sphaleron processes can be understood as transitions between different vacuum configurations of the electroweak theory, characterized by integer Chern-Simons number $N_{CS}$. In each transition, $B$ and $L$ change by a number determined by electroweak anomalies; in the SM, one has $\Delta B=\Delta L=3$ for a transition with $\Delta N_{CS}=1$, which can be captured with an effective 12-fermion interaction. The sphaleron itself is a meta-stable classical configuration at the top of the minimal energy barrier between the different vacua; the barrier height is  $E_{\rm sph}=9.1$ TeV. In the vacuum and at low energies, transitions proceed through quantum tunneling, while within a thermal plasma with temperatures above $E_{\rm sph}$ the rate becomes unsuppressed. It has been speculated that if high energy colliders were to reach center-of-mass energies above $E_{\rm sph}$, $B+L$-violating sphaleron processes could be observable. Naively, an energy above the barrier height should allow for unsuppressed transitions. However, the field path with a minimal energy barrier passes through the sphaleron, which as a classical configuration can be thought of as a coherent superposition of quantum excitations. Then one expects a ``few-to-many'' suppression factor from the transition from a two-particle initial state to the intermediate sphaleron configurations. 

To elucidate which effect dominates, different calculations have been attempted, reviewed for example in \cite{Mattis:1991bj,Tinyakov:1992dr,Guida:1993qy,Rubakov:1996vz}. In principle, tunneling problems in gauge theories can be dealt with a semiclassical expansion (instanton perturbation theory \cite{Belavin:1975fg,tHooft:1976snw}), and it was initially found in refs.~\cite{Ringwald:1989ee,Espinosa:1989qn} that, when accounting for the effect of gauge boson emission --the number of gauge bosons is not restricted by anomalies, as opposed to the case of fermions-- the cross-section grew exponentially with energy. It was subsequently confirmed that higher-order contributions from boson emission to the cross-section do indeed resum to an exponential of a function of the energy --the ``Holy Grail function''-- which however cannot be calculated reliably with instanton perturbation theory for $E\gtrsim E_{\rm sph}$, precisely the energy range of interest in which the tunneling suppression could be overcome \cite{Arnold:1990va,Khlebnikov:1990ue,Arnold:1991cx,Mueller:1991fa,Khoze:1990bm}. Thus one has to look for alternative techniques, and there have been calculations using dispersion relations \cite{Zakharov:1990dj,Porrati:1990rk,Khoze:1990bm,Khoze:1991mx,Ringwald:2002sw}, direct semiclassical evaluations of the total cross-section  \cite{Rubakov:1991fb,Tinyakov:1991fn,Bezrukov:2003er,Bezrukov:2003qm}, and estimates based on the dynamics of $N_{CS}$ in its periodic potential \cite{Tye:2015tva,Tye:2017hfv}. The different answers differ by many orders of magnitude and the issue of observability of these processes is still under debate.

In the work presented at the conference \cite{Cerdeno:2018dqk}, we chose not to focus in the exponential contributions to the cross-section coming from gauge boson emission, centering instead on the effect of exotic fermions. Fermions affect non-exponential contributions to the cross-sections, which can however still be large. The number of fermions in each sphaleron transition is constrained by anomalies, and as such there is no analogue of the enhancement coming from summing over all possible bosonic final states; due to this, we expect that leading-order instanton calculus should accurately capture the fermionic contributions. Our estimate resembles the one done in the Standard Model in ref.~\cite{Ringwald:1989ee}, yet our treatment is original when it comes to correcting the instanton density in order to account for decoupling effects. 

The rest of the paper is organized as follows. In section \ref{sec:vertices} we summarize how chiral $SU(2)_L$ anomalies constrain the number of fermions that participate in the $B+L$-violating interactions. Next, section \ref{sec:Lagrangian} summarizes the calculation of the effective interaction vertices using instanton techniques. Estimates of the impact of exotic fermions are given in section \ref{sec:results} for $B+L$-violating reactions involving a new Dirac fermion in the fundamental of $SU(2)_L$, a Weyl spinor in the adjoint, and both of the former. These new particles charged under $SU(2)_L$ are the same as the ones present in  the Minimal Supersymmetric Standard Model (MSSM), or in related variations such as the Split SUSY scenario. The Dirac fermion and the Weyl spinor correspond to the two Weyl Higgsinos and the $SU(2)_L$ gaugino, respectively. 

\section{\label{sec:vertices} Chiral anomalies and effective interaction vertices}

\begin{figure}
\centering
\raisebox{.5cm}{\includegraphics[width=0.25\textwidth]{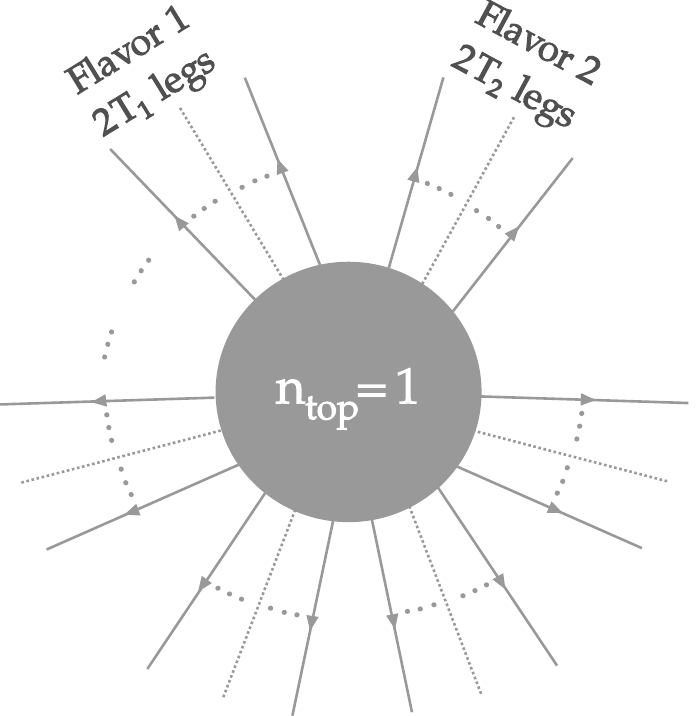}}
\includegraphics[width=0.29\textwidth]{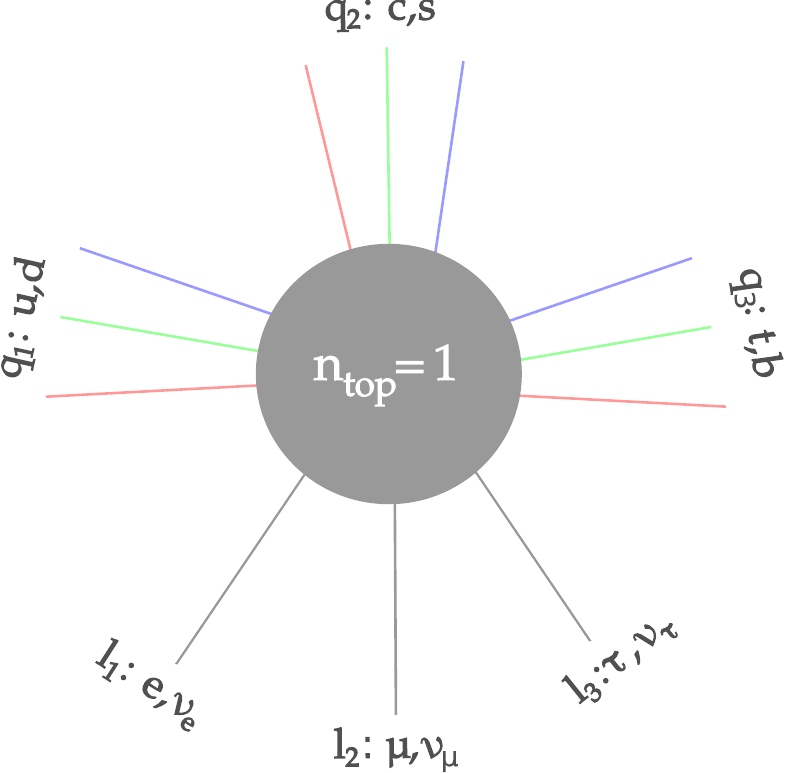}
 \caption{\label{fig:vertices}Anomalous interaction vertices in an $n_{\rm top}=1$ background for a generic theory (left), and for the SM (right).}
\end{figure}

In a theory with massless Weyl spinors $\psi_k$ charged in representations $r_k$ (which we will call ``flavors'') of $SU(2)_L$ with Dynkin index $T_k$,
\begin{align}
\label{eq:TR}
\text{Tr}_{r_k}\, T^a T^b=T_k\,\delta^{ab},
\end{align}
quantum effects imply that  global symmetries of the form $S:\psi_k\rightarrow e^{iq^S_k\alpha} \psi_k$ are anomalous, so that their associated charges $Q_S$ are no longer conserved but satisfy instead \cite{Adler:1969gk,Bell:1969ts,Bardeen:1969md}
\begin{align}
\label{eq:charge}
Q_S(t=\infty)-Q_S(t=-\infty)=\int d^4 x\,\partial_\mu J_S^\mu = N_S n_{\rm top}\ , \quad N_S=\sum_k q^S_k n_{\rm top},
\end{align}
where $n_{\rm top}$ is the integer topological charge of a given gauge-field background. By considering simple symmetries $S_k$ corresponding to the Weyl fermion number for the $k$-th flavor, i.e.  $q^{S_k}_m= \delta_{mk}$, it follows that for every Weyl spinor in a representation $r_k$, the minimal amount of violation of its Weyl fermion number is $2T_k$ units. Hence, the anomalies predict effective interactions involving $2T_k$ copies of each fermion charged under $SU(2)$ (see fig.~\ref{fig:vertices}). Moreover, such interactions are associated with gauge field backgrounds of unit topological charge. In the Standard Model, with fermions in the fundamental of $SU(2)_L$ having Dynkin index $T_{F}=1/2$, one predicts vertices involving 3 copies of each quark doublet (the factor of 3 accounting for the $SU(3)$ colour multiplicity) and one copy of each lepton doublet. This gives a 12-fermion interaction vertex, which violates $B$ and $L$ by 3 units, as also illustrated in fig. ~\ref{fig:vertices}. In a theory with an exotic Dirac fermion in the fundamental of $SU(2)_L$ (equivalent to two fundamental Weyl spinors), one predicts vertices with 14 fermions, involving two exotic legs, while for the SM plus a Weyl spinor in the adjoint, with $T_A=2$, one expects 16-fermion interactions with 4 exotic legs. For SUSY-like scenarios with Higgsinos in the fundamental and an electroweak gaugino in the adjoint, anomalies predict 18-fermion vertices.
If the exotic fermions are massive, their corresponding symmetries $S_k$ are explicitly broken by the mass terms. However, one can treat the masses as spurions with a Weyl fermion number of -2, so that the mass terms ${\cal L}\supset -M_k\,\psi_k \psi_k+c.c.$ remain formally invariant. Then the previous conclusions about the violation of the chiral charge by the anomaly still apply, but now allowing for replacing pairs of massive Weyl spinors with  insertions of $M^*$, as both have a Weyl fermion number of +2.\footnote{Note that adding pairs of spinors and dividing by $M^*$ is forbidden by the requirement of a well-defined massless limit.} In particular, with enough insertions of masses one recovers SM-like interactions, as expected from the decoupling theorem: at low enough energies, the SM should arise as an effective theory. Finally, it should be noted that the previous vertices can in principle be dressed with arbitrary number of Higgs and gauge bosons, as their numbers are not constrained by the anomalies.
\begin{figure}
\centering
\includegraphics[width=0.27\textwidth]{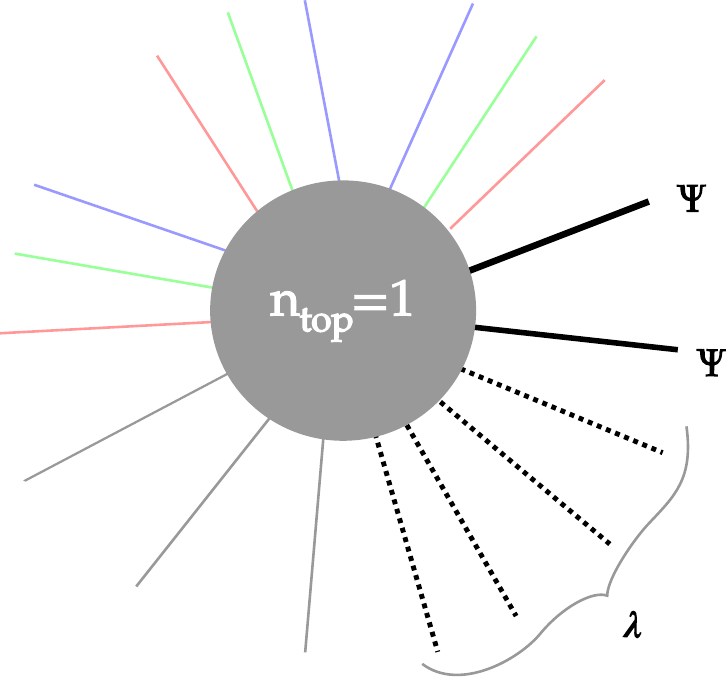}%
\includegraphics[width=0.27\textwidth]{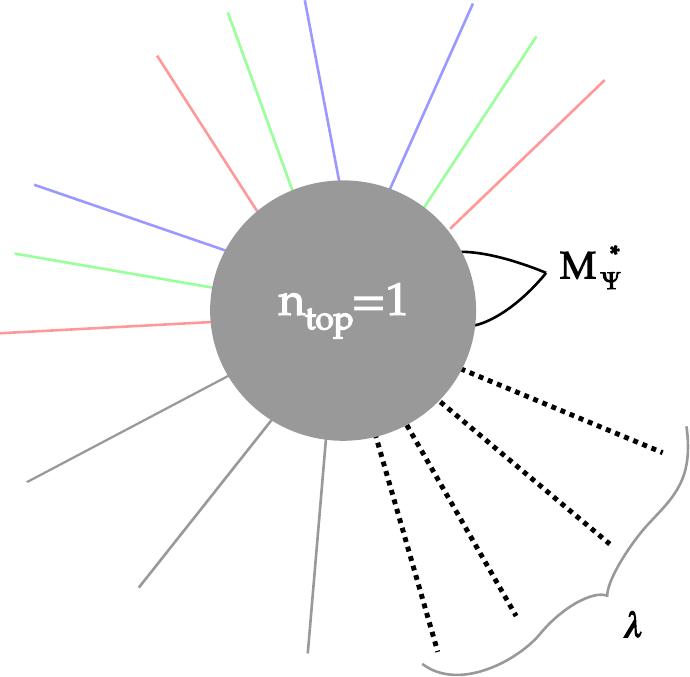}%
\includegraphics[width=0.27\textwidth]{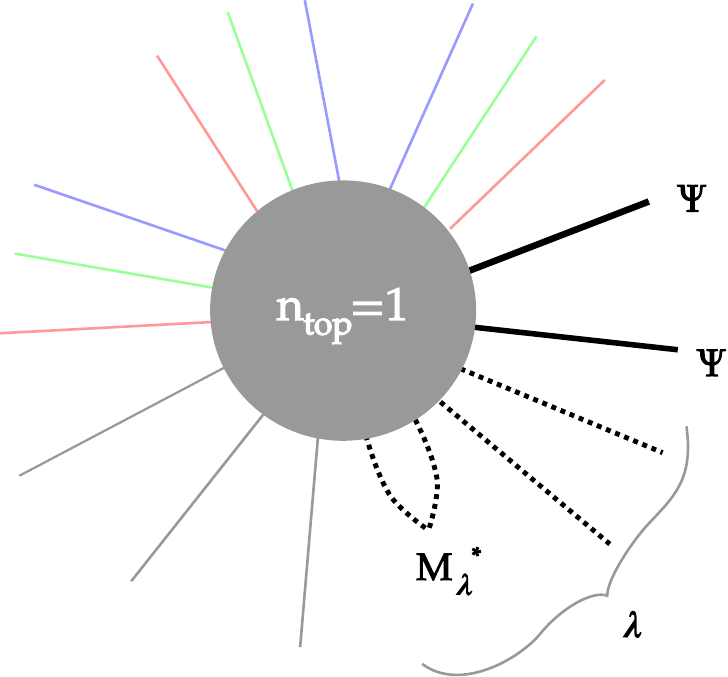}
\includegraphics[width=0.27\textwidth]{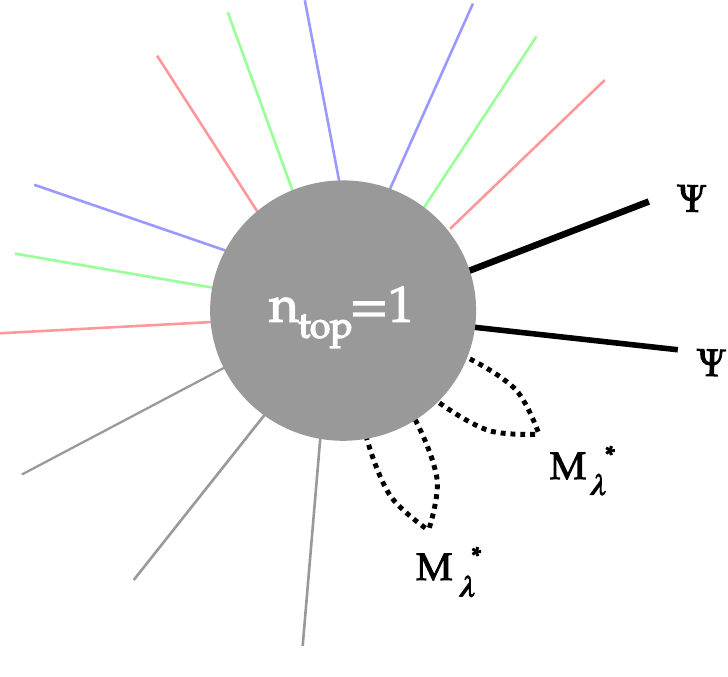}%
\includegraphics[width=0.27\textwidth]{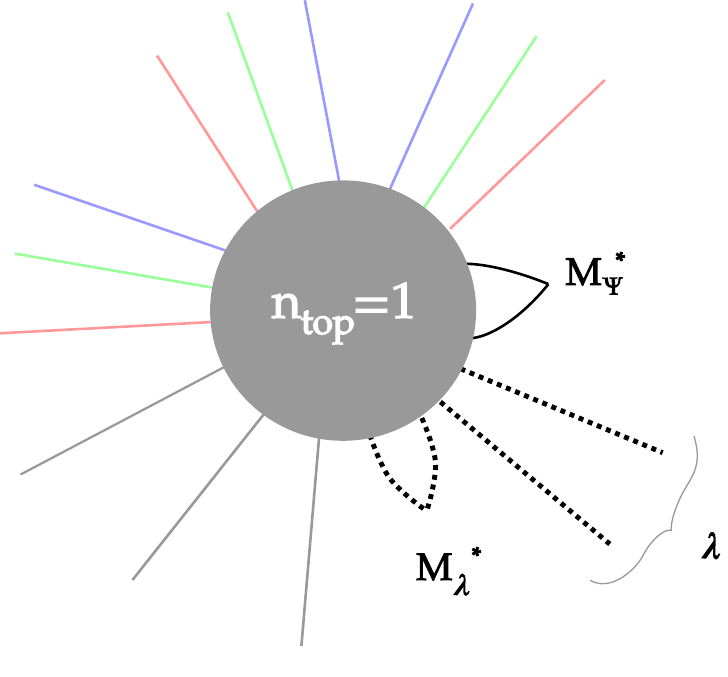}%
\includegraphics[width=0.27\textwidth]{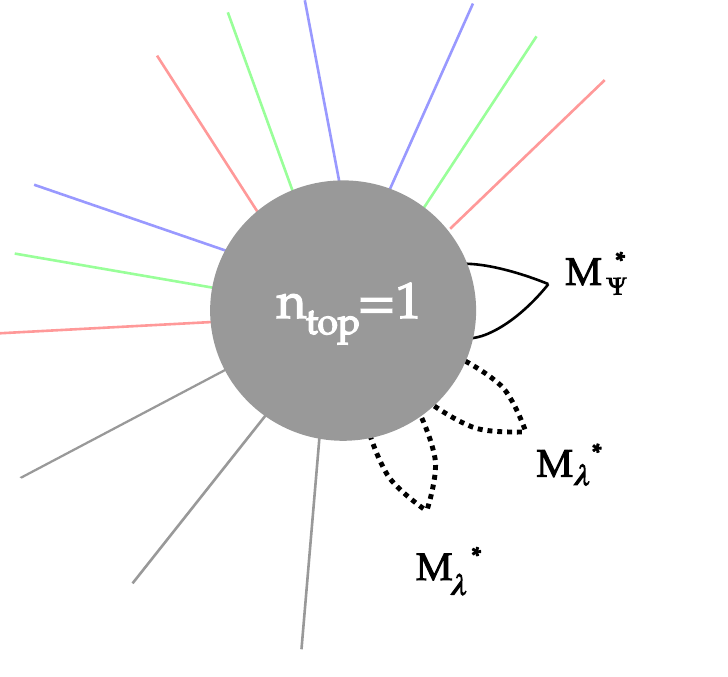}
 \caption{\label{fig:vertices2}Anomalous interaction vertices in theories with the same $SU(2)_L$-charged fermions as in the MSSM. Dark solid lines correspond to Weyl spinors $\Psi$ in the fundamental of $SU(2)_L$, while dotted dark lines denote spinors $\lambda$ in the adjoint. Note the inclusion of mass insertions, and the SM-like vertex in the lower right.}
\end{figure}
\section{\label{sec:Lagrangian} Effective instanton Lagrangians and decoupling effects}
We have constructed effective Lagrangians for the anomalous vertices by using leading-order instanton computations. The effective Lagrangians are such that their associated amputated Green functions --as needed for scattering-- match the corresponding Green functions in the instanton background, amputated with ordinary propagators \cite{tHooft:1976snw,tHooft:1976rip,Shifman:1979uw}. In our calculations, we ignore details of the spinor and group contractions within the Green functions, as we are interested in order-of-magnitude estimates. The effective Lagrangians involve an integration over the scale $\rho$ of the semiclassical instanton solutions, weighed by an instanton density function and $\rho$-dependent fermionic form-factors. In order to remain compatible with decoupling --as is crucial for recovering the SM result from the BSM theories-- we implement a modification of the instanton density after crossing a heavy mass threshold $|M|$. The modifications involve powers of $\rho |M|$, and it is a nontrivial fact that such insensitivity to the phases of the mass matrices actually follows from a correct matching of the $SU(2)$ $\theta$ angle across a threshold. The ensuing effective Lagrangians, including the effect of gauge boson emission, have the following form:
\begin{equation}
\label{eq:Lschematic3}\begin{aligned}
{\Delta L}\supset&\sum_{n_W,n_h} \int\frac{d\rho}{\rho^5}\,\tilde C_I(\rho)\,(-\sqrt{2}\pi^2 \rho^2 v h)^{n_h}\left(-\frac{4\pi^2\rho^2}{g}\eta_{a\mu\nu}\partial_\nu W^{a}_{\mu}\right)^{n_W}\\
 &\times\prod_{[k,l]}\left\{(\rho|M_{kl}|)^{N^0_{kl}b_{kl}}\sum_{j=0}^{N^0_{kl}} \left({\cal F}_{kl}\right)^{j}(\psi_k\psi_l)^{j}(\rho M^*_{kl})^{N^0_{kl}-j}\right\}\\
&\times\prod_{m}\left\{(\rho|M_{mm}|)^{1/2N^0_{mm}b_{mm}}\sum_{i=0}^{1/2N^0_{mm}} \left({\cal F}_{mm}\right)^{i}(\psi_m\psi_m)^{i}(\rho M^*_{mm})^{1/2N^0_{mm}-i}\right\},\\
b_{m n}=&\,\left\{\begin{array}{cc}
           0, &\rho |M_{m n}|<1,\\
           -1/3, &\rho |M_{m n}|\gtrsim1. 
          \end{array}\right.
\end{aligned}\end{equation}
Above, $\tilde C_I(\rho)$ is the instanton density, depending on the $SU(2)$ running coupling and the Higgs VEV. $n_W, n_h$ correspond to the number of Higgses and W bosons, while $k,l,m$ label representations of Weyl fermions. In particular, $[kl]$ denotes pairs $(k,l)$ with $k\neq l$, associated with Weyl spinors in the fundamental representation and coupled by a mass term $M_{kl}$, so that they can be grouped into a massive Dirac spinor. On the other hand, the index $m$ labels Weyl spinors in the adjoint representation, which can be related to 4-component Majorana spinors with masses $M_{mm}$. ${\cal F}_{pq}$ denote fermionic form factors that depend on $\rho$ and the representations of the fermions labelled by $p,q$, while $N^0_{pq}=T_p+T_q$, with $T_q$ the Dynkin index of eq.~\eqref{eq:TR}. The powers of $\rho |M_{pq}|$ with the exponents $b_{mn}$ changing across thresholds implement the decoupling corrections. We refer to the paper \cite{Cerdeno:2018dqk} for more details.
\section{\label{sec:results}Estimates of fermion enhancement}
\begin{figure}
\centering
\includegraphics[width=0.4235\textwidth]{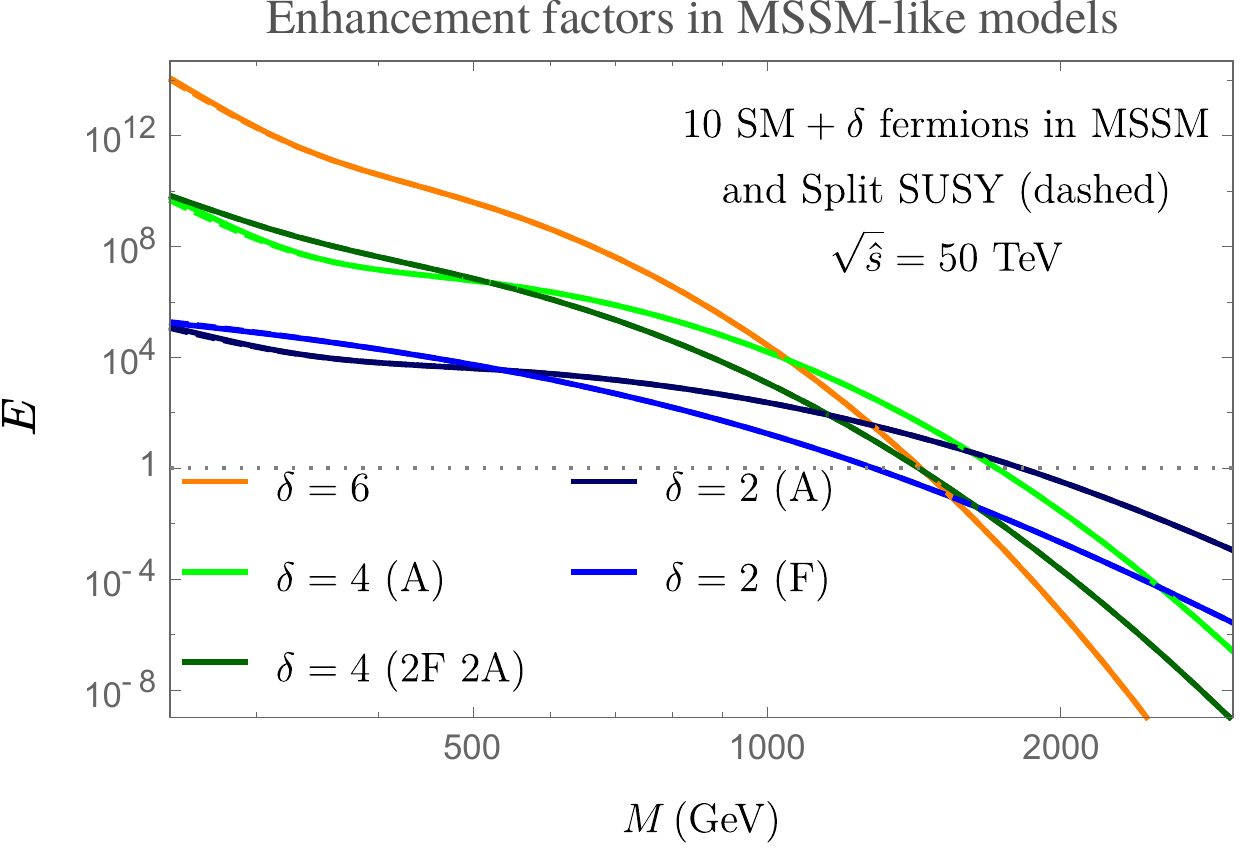}%
\includegraphics[width=0.5765\textwidth]{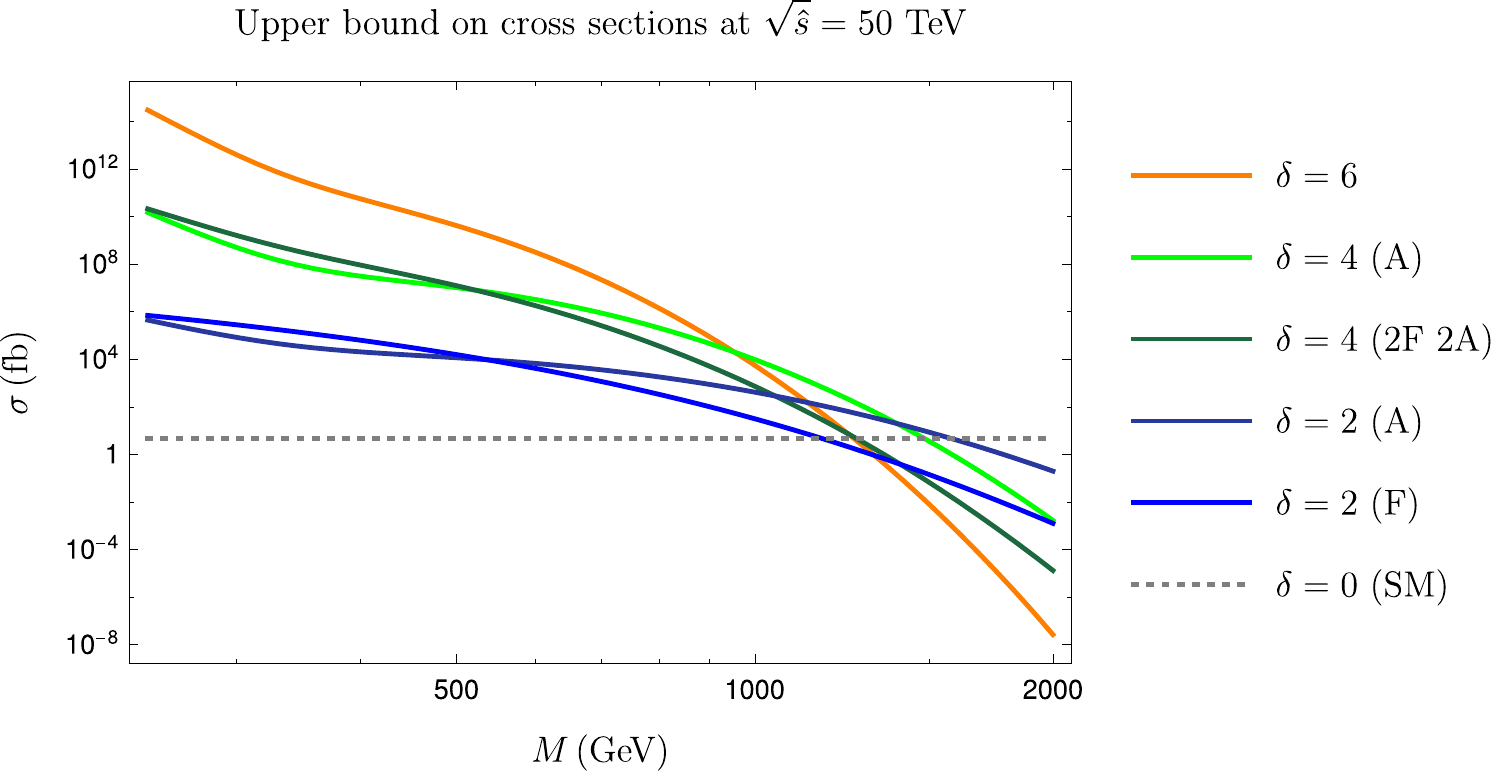}%
 \caption{\label{fig:results} Left: enhancement factors in MSSM-like models, for interactions involving  $\delta$ exotic fermions in the fundamental (F) and adjoint (A). Right: Upper bounds for cross sections.}
\end{figure}

To estimate the impact of BSM fermions, we use the effective Lagrangian of eq.~\eqref{eq:Lschematic3} to compute ratios of cross-sections for processes involving exotic fermions over cross-sections for SM-like processes. We consider two-quark initial states with a fixed center-of-mass-energy $\sqrt{\hat{s}}$. By taking ratios, we eliminate the uncertainty in the effect of gauge boson emission. The ratios define enhancement factors $E(\sqrt{\hat{s}},n_i,M_i)$, where $M_i$ are the heavy fermion masses, and $n_i$ their multiplicity in the interaction vertex. For simplicity we will assume degenerate heavy fermions, $M_i=M$, for which the enhancement factors only depend on $M$ and the  total number of exotic fermions in the interaction vertex, $\delta=\sum_i n_i$. Despite the fact that the above instanton Lagrangian only captures  boson emission accurately for $\sqrt{\hat{s}}<E_{\rm sph}$ --as it only accounts for leading term in an expansion of the Holy Grail function in $\sqrt{\hat{s}}/E_{\rm sph}$-- we have checked that the effect of the new fermions in the  boson emission can be captured by a simple modification of the SM result, in which the SM factor accounting for the boson production is to be evaluated not at $\sqrt{\hat{s}}$ but at the maximum energy available for the bosons, i.e. at $\sqrt{\hat{s}}-\delta M$. We conjecture that this will hold for the full Holy-Grail function at higher energies. Thus, using the SM base rate in ref.~\cite{Ringwald:2003ns}, our enhancement factors and the above ansatz for the effect on gauge boson emission, we can estimate cross-sections as 
\begin{align}
 \label{eq:sigmaF2}
\sigma_{B+L}^{2\rightarrow {\rm any}}=\frac{ E(\sqrt{\hat s},\delta,M)}{m^2_W}\left(\frac{2\pi}{\alpha_W}\right)^{7/2}e^{-\frac{4\pi}{\alpha_W}F[(\sqrt{\hat{s}}-\delta M)/E_{0}]}.
\end{align}
For actual cross-section calculations, we need an estimate of the Holy Grail function $F$, for which we choose the lower bound given in \cite{Bezrukov:2003er,Bezrukov:2003qm}, which gives an upper bound for the cross-sections. We show in fig.~\ref{fig:results} results for ratios of cross sections in MSSM-like scenarios for $\sqrt{s}=50$ TeV, as well as total cross-section estimates using formula \eqref{eq:sigmaF2}.

\section{\label{sec:conclusions}Conclusions}

We have analyzed the effect that massive exotic fermions charged under $SU(2)_L$ can have in nonperturbative $B+L$-violating processes. Chiral anomalies predict that the former interactions can be accompanied by the emission of exotic fermions, and we set out to elucidate whether the rates of the exotic $B+L$-violating reactions at colliders can dominate over those that only involve the Standard Model particles. Using instanton techniques, we have computed enhancement factors associated with the new fermions, and found that they can have a very large impact if the new particles are light enough. Since the enhancement factors are computed by taking ratios of rates, we expect the results to be independent of the uncertainties associated with the effect of boson emission. Regarding the latter, we performed numerical calculations suggesting that the effect of fermion emission can be simply captured 
by a change in the argument of the Holy Grail function. This allowed us to compute upper bounds on partonic-level cross-sections, building up on Standard Model results in the literature. The large enhancement factors due to the exotic fermions suggest that if $B+L$-violating interactions were ever detected at a high-energy experiment, they could be tied to new physics.

\section*{Akcnowledgements}
D.G.C. is supported by the SFTC. P.R. is funded by the Graduiertenkolleg \textit{‘Particle physics beyond the Standard Model’} (GRK 1940). The work of K.S. was partially supported by the National Science Centre, Poland, under research grants DEC-2014/15/B/ST2/02157 and DEC-2015/18/M/ST2/00054, as well as by MEXT KAKENHI Grant-in-Aid for Scientific Research on Innovative Areas, Japan, Grant Number JP16K21730.  C.T. acknowledges support from the Collaborative Research Centre SFB1258 of the Deutsche Forschungsgemeinschaft (DFG). 
\section*{References}
\bibliographystyle{iopart-num}
\bibliography{library}

\end{document}